# Green Bitcoin: Global Sound Money

Heung-No Lee, Young-Sik Kim, Dilbag Singh, and Manjit Kaur

**Abstract**

Modern societies have adopted government-issued fiat currencies many of which exist today mainly in the form of digits in credit and bank accounts. Fiat currencies are controlled by central banks for economic stimulation and stabilization. Boom-and-bust cycles are created. The volatility of the cycle has become increasingly extreme. Social inequality due to the concentration of wealth is prevalent worldwide. As such, restoring sound money, which provides stored value over time, has become a pressing issue. Currently, cryptocurrencies such as Bitcoin are in their infancy and may someday qualify as sound money. Bitcoin today is considered as a digital asset for storing value. But Bitcoin has problems. The first issue of the current Bitcoin network is its high energy consumption consensus mechanism. The second is the cryptographic primitives which are unsafe against post-quantum (PQ) attacks. We aim to propose Green Bitcoin which addresses both issues. To save energy in consensus mechanism, we introduce a post-quantum secure (self-election) verifiable coin-toss function and novel PQ secure proof-of-computation primitives. It is expected to reduce the rate of energy consumption more than 90 percent of the current Bitcoin network. The elliptic curve cryptography will be replaced with PQ-safe versions. The Green Bitcoin protocol will help Bitcoin evolve into a post-quantum secure network. In addition, it improves the properties of Bitcoin's hash PoW while addressing environmental concerns.

## I. INTRODUCTION

WE the global citizens not by our choice live in a world of boom-and-bust cycles created by the Federal Reserve Board (FED) of the United States. In the boom phase of a cycle, the FED supplies debt-monetized US dollars to the world, and supplied USDs are used to purchase real items, goods, and services from developing countries. Such supplies prop up bubble markets, such as the US housing, stock, and derivatives markets. Global elite financial institutions, such as investment banks and hedge funds, benefit the most from making risky investments. When the bust part of the cycle comes, the working class worldwide suffers from

This work was supported in part by a National Research Foundation of Korea (NRF) grant funded by the Korean government (MSIP) (NRF-2021R1A2B5B03002118) and the Ministry of Science and ICT (MSIT), Korea, under the Information Technology Research Center (ITRC) support program (IITP-2021-0-01835). This work was also supported by the Tech Incubator Program for Startup (TIPS) program (S3306777) awarded in June 2022.

Heung-No Lee is with Gwangju Institute of Science and Technology, Gwangju, 61005, South Korea and affiliated with LiberVance,Co., Ltd., Gwangju, South, Korea.
Dilbag Singh and Manjit Kaur are with Gwangju Institute of Science and Technology, Gwangju, 61005, South Korea.
Young-Sik Kim is with Chosun University, Gwangju, 61452, South Korea, and affiliated with LiberVance, Co., Ltd., Gwangju, 61005, South Korea.

Corresponding author: Heung-No Lee (e-mail: heungno@gist.ac.kr).

inflation and market crashes. In the aftermath of a market crash, governments bail out financial institutions to prevent chain bankruptcies. Such centralized planning has disrupted societies globally [21][29]. The boom-and-bust volatility peaked with the Dow to gold ratio has become increasingly extreme [6]. Inequality is a prevalent condition worldwide. Work ethics fade. Growing are speculative markets. Keeping this system in its current form does not help advance humanity to the next level.

Throughout history, money has taken many forms, including gold, silver, copper, salt, and seashells [21]. As modern society developed, government-issued fiat currencies became normal, many of which exist in the form of digits in banks and credit accounts today. The soundness of money is determined by its stability; if it is stable, it can function as a medium of exchange, unit of account, and store of value. Central banks know this and seek to stabilize their currency by controlling money supply as the economic condition changes. Today, however, this flexibility is often misused and overused, especially by a new government that needs to satisfy voters and make way for its political agenda. The boom comes from stimulation, and the bust comes back from monetary tightening. It seems clear that innovative measures are needed to address the current situation, which depends entirely on central bank policy decisions.

Today, central banks are often tightly coupled with political powers. When new presidents come into the office, they are tempted to use the central bank's power in money creation capability to fulfill their political agenda. The stability of money, and thus the market order, is broken in the name of economic growth and stabilization. Owing to the manipulation of money, monetized debts, inflation of asset prices, and growing inequality are rampant [4]. Thus, there is a need to restore sound money whose supply is independent of governments' control, which is left alone to a free market and its self-regulation mechanisms.

Bitcoin was humanity's first success in creating decentralized money [21][41]. It realized a sound money Fredrick Hayek proposed in his book, *The Denationalization of Money* [22]. Its success was possible because of cryptographic technologies, such as SHA, digital signature algorithms, and elliptic curve cryptography. Currently, Bitcoin is one of the soundest currencies, but it still does not fully satisfy itself as sound money. For example, it does not work as a means of payment, as blockchain networks incur transaction costs that are too high for daily spending. A lightning protocol (along with other second-layer solutions) [53] facilitates off-chain transactions at high speeds. Unfortunately, these solutions are not secure against post-quantum (PQ) attacks.

These cryptographic primitives must be upgraded regularly. Otherwise, their use is limited and useless if not upgraded. IBM announced its plan to introduce quantum computers with more than four thousand qubits by 2023 [26]. With such advances, cryptographic algorithms used in Bitcoin are on the verge of breaking.

The main contributions of this paper are as follows. This paper reviews Hayek's sound money and discusses the pressing need to restore it in our society. The advantages of Bitcoin concerning its robust performance, such as simplicity in consensus and time-energy-borne wealth characteristics, are emphasized. Subsequently, a post-quantum (PQ) safe Green Bitcoin protocol is proposed. The proposed protocol can help achieve the majority of sound money properties. Green Bitcoin comprises two major parts: a PQ secure verifiable (self-election) coin-toss (VCT) function and a novel PQ secure proof-of-computation (PoC) primitive. The PoC part is built based on a newly published finding known as the error-correction code anti-ASIC proof-of-work (ECCPoW) [30][31][33][45]. PoC primitives will make PQ safer than the ECCPoW. Environmental concerns can be addressed with the VCT function, aided by the PoolResistantComp protocol. This can be used to control the network's energy consumption efficiency. Critical components of opcodes, such as digital signature algorithm, will be enhanced, and elliptic curve cryptography will be replaced with PQ safe versions. The Green Bitcoin protocol will help Bitcoin evolve into a post-quantum secure network. In addition, it helps achieve good properties of Bitcoin's hash PoW while alleviating environmental concerns.

The remainder of this paper is organized as follows: Section II discusses the significance of cryptocurrency and blockchain. Section III discusses the benefits of Bitcoin, sound money, and problems. Section IV presents the Green Bitcoin proposal. Novel PQ secure primitives are discussed in Section V, and a Green Bitcoin testbed and discussion are presented in Section VI. Section VII concludes this paper.

## II. CRYPTOCURRENCY AND BLOCKCHAIN

Cryptocurrencies are digital currencies created through encryption algorithms that can be used as a form of payment. Using encryption technologies, cryptocurrencies can serve as currency and virtual accounting systems. Digital wallets are used for cryptocurrency trades. Blockchain networks were developed to provide decentralized requirements for cryptocurrencies. Consensus, virtual machines, and peer-to-peer (p2p) networking are the three primary components of a blockchain. One of the most pressing demands is to 1) update the consensus mechanism that allows a new PQ secure and decentralized blockchain network and 2) upgrade the cryptographic primitives used in consensus and virtual machines to be PQ safe.

Currently, blockchains are not PQ secure, there are environmental concerns and scalability issues. Several major projects have implemented Proof of Stake (PoS) paired with a Byzantine agreement (BA) algorithm to resolve these concerns. The PoS and BA algorithms are perhaps good for a fast-computing platform, but they are not secure enough for a global monetary-grade blockchain. The ideas are not new; the BA algorithm (developed in the 1980s [12][32]) relies on communication across committee nodes to reach a consensus, rendering it subject to various assaults, including DDoS (distributed denial of service) and network partition attacks. To scale it well in terms of the number of nodes, the committee size cannot but remain small; hence, decentralization is compromised.

Blockchains are a dear but expensive solution. The blocks are stored redundantly within every consensus-participating node, and all nodes perform the same work. Each new block is made with an effort, an enormous amount of time and energy, by the entire network. This is the source of the immutability of the records stored in the blockchain; the network must be decentralized to the maximum possible extent. The greater the number of individual nodes participating, the more secure the network is regarding censorship resistance, thwarting Sybil, and double-spending attacks [27][41]. Consider Bitcoin; each block contains a massive amount of computational energy stored in each block. If the block needs to be forged again, it requires the same amount of energy to be stored. On the one hand, the large redundancy and numerous independently working nodes doing the same work can be viewed as a source of security. On the other hand, it can be viewed as a waste of resources and a waste of energy. The blockchain trilemma [5][7][14]—it is difficult to achieve the three blockchain properties simultaneously such as scalability, decentralization, and security—represents a scalability challenge caused by inefficient resource consumption. The complaint about energy issues leads to environmental concerns.

Blockchain technology requires a simple protocol to withstand attacks and perform robustly for years to come [33]. The consensus mechanism should be able to maximize the resilience of numerous unknown attack vectors. How can we make it simple while accommodating many p2p nodes working together to reach a consensus? The participating nodes must make timely judgments and choose one block from several candidate blocks to be the new block attached to the status quo chain. A timely consensus decision should be distributed among numerous independent working nodes over the internet. Consequently, an agreement must be reached with as few contacts as possible among the p2p nodes.

Frequent network delays and partitions occur on the internet. They might be caused by momentary router failures, traffic congestion, or purposeful antagonistic activities. Therefore, a global blockchain network must be resilient against possible attacks and losses.

A consensus mechanism using hash function-based PoW [41] has been shown to provide the most decentralized and secure operations. Thus, it is challenging to design a consensus mechanism that achieves all these needs: a large number of p2p nodes, making timely decisions while working independently, with minimal inter-node communication.

## III. BITCOIN, SOUND MONEY, PROBLEMS

### A. Bitcoin

Bitcoin [41] is a newly developed type of money known as "cryptocurrency." Some in the Bitcoin community believe that Bitcoin already represents sound money [23][24]. Sound money is defined as money that has a purchasing power determined by markets, independent of governments and political parties. There would be a vast difference between a world with Bitcoin and one without Bitcoin. While governments continue to print cheap money, ordinary people can resort to Bitcoin. Thus, the wealth gap can be significantly reduced. We will come closer to a produce-first-and-spend economy rather than a debt-monetized spending-first economy. Bitcoin encourages savings and autonomous management of money; it does not allow bailouts. Governments are not held accountable for policy failure. Tax-paying individuals and corporations are the entities who are held accountable for the economic setbacks caused by policy failures. The use of the power to mint the currency held by governments (non-responsible entities) should be limited. That way, central banks will no longer be able to punish diligent savers by sprinkling them with debt-based currencies. With a stable store of value, people can confidently plan for the future. Zombie companies and bubble markets that thrive on government budgets and easy monetary policies will find it difficult to survive.

An organism's code of life is written at its conception. Bitcoin's DNA, or genetic code, was carefully crafted by Satoshi as the soundest money ever created. Bitcoin's genetic code can be considered a set of instructions designed to incentivize the coordination and organization of cellular functions.

The genetic code of Bitcoin [41] is as follows:

- Bitcoin had to be ignited to become real, so Satoshi coded a fixed supply (21 million Bitcoins) into its DNA. Sound money comes from this fixed supply. Due to the increase in users, miners and developers, Bitcoin has become more expensive over time. Thus, the feedback loop has become self-reinforcing.
- The mining function, that is, the one utilizing a hash PoW, is the metabolism and defense mechanism of Bitcoin. For instance, Bitcoin consumes a lot of electric energy to make new blocks and create virtual walls to protect the network from hackers. The anti-fragility of Bitcoin is attributed to PoW, which makes it more resistant to attacks as it grows.
- The Bitcoin network generates a new block every 10 min on average. This rate is carefully determined for the robust operation of the nodes. Bitcoin nodes are scattered all over the globe; thus, they are separated over long distances. However, they can still communicate and coordinate effectively and have never stopped making new blocks since their inception in January 2009.

### B. What Is Sound Money?

The 1974 Nobel Prize in Economics Laureate Friedrich Hayek wrote the *Denationalization of Money* (1976) [22]. He said privately issued money, which continues to develop through competition, will inevitably be superior to fiat money, which does not evolve because the government has exclusive power to supply it.

> "I am more convinced than ever that if we ever again have sound money, it will not come from the government. This is issued by a private enterprise,"　　　　　　　　　　　　　　　　　　　　Friedrich Hayek, 1977

Hayek emphasizes sound money in his book. Literally translated, it can be interpreted as honest money. However, one may wonder what sound money is and why he emphasized it. What will happen if money is sound, and what will happen if it is not? As a firm definition is not given in his book, the answers to these questions can be examined via the statements he provides in his book.

Today, we live in the age of fiat currencies. For example, the US dollar is a fiat currency. Fiat means "by government decree." By law, the U.S. government declares the dollar to be a currency. Because the United States has the most powerful military and the largest economy in the world, it has gained worldwide trust, and the US dollar has become the world reserve currency.

Individuals earn sound money through solid value-creation processes. To produce more gold, one has to mine deeper in the gold mines. Gold has been utilized for thousands of years as a means to store value for future use. No government can print a sound money. No government can devalue the money and weaken the savings. To earn other people's money, one must provide equivalent time and energy for others. Governments should not be exceptions. They must also be prudent and considerate. They must win the hearts of their peoples with reliable, agile and considerate services, and a clear vision. They shall be limited to printing new money.

### C. What Is the Problem If Money Is Not Sound?

The US dollar has caused serious social problems, as revealed by several studies [23][25]. The dollar has been overused by the US government and the central bank FED in the name of "stimulating the economy," "resolving the financial crisis," and "fighting the pandemic." Whenever a bust phase occurred, dollars were printed and supplied to society. Recent examples include quantitative easing measures to address the financial crisis caused by subprime mortgages in 2008. The US government issues treasury bonds and the FED provides dollars to society by buying them. The newly issued dollars spread to the world society. Major Western advanced countries will also have to expand the supply of their currencies to resolve the negative side effects of the large-scale supply of dollars.

This creates a situation where debt increases in major countries worldwide, and an oversupply of money is widespread. The additional money supplied each year causes real estate prices in major cities to skyrocket. The value of financial assets, such as the S&P 500 ETF in the US, continues to grow in a right-upward way by more than 15% annually. There is a huge wealth gap between those who can own these assets and those who cannot. Speculative markets in which unearned income can be obtained as much as the amount of over-supplied fiat currency will grow. However, the value of honest work based on diligence, knowledge, and skills decreases. Knowledge-seeking and corporate activities weaken. What labor activity, what knowledge-seeking activity, or what company's business model could produce growth rates higher than that of the S&P 500?

In a society where honest money reigns, bubbles weaken, and honest work is promoted. Companies invest in new knowledge- and skill-creation projects and develop new products and services. Honest labor and knowledge-seeking activities are vibrantly carried out, the workforce participating in production activities increases, and the economy grows robustly. Because there is no oversupply of money, no speculative bubble markets grow. As the speculative market disappears, more people can find a stable life through honest work.

Gold has been used as honest money for thousands of years [21][22]. Honest money is very important to humans because it performs useful roles such as a means of exchange, a measure of value, and store of value. However, gold is not easily produced. Producing even a single gram of new gold requires someone to work very hard over a long period. The only way to produce gold is to mine it from the ground. All gold that exists on the earth's surface has already been mined. Therefore, to mine gold today, you have to dig deeper. However, the production of gold has become increasingly difficult. This rarity is a key characteristic of gold. Because of these properties, gold was used as sound money.

The United States adopted the gold standard through the Bretton Woods Agreement of 1944. An ounce of gold was pegged at 35 US Dollars. Currencies in the Western world, such as the Franc (France), Pound (United Kingdom), Mark(Germany), and Yen(Japan), were pegged to the US Dollar, and currencies in developed countries were pegged to gold. Therefore, these currencies of major countries around the world was honest. On August 15, 1971, the 37th President of the United States, Nixon, announced [55] that he would temporarily withdraw from the gold standard, shocking the world. Since then, there have been many reports that easy monetary policy has created a polarization of wealth [25].

### D.  Is Bitcoin Sound Money?

Bitcoin has already acquired the status of an asset (a means of saving and investing) in many countries. El Salvador became the first country in the world to recognize Bitcoin as its legal tender on September 7, 2021. The US Securities and Exchange Commission (SEC) approved the Bitcoin Futures ETF on October 16, 2021 [9].

Is Bitcoin already the sound money Hayek was talking about? This is believed to be so among many Bitcoin enthusiasts [23][24]. According to Hayek's proposition of sound money, however, it should evolve through competition. Therefore, many Bitcoin hard forks should emerge in the future and continue competing to provide better services. As numerous new Bitcoins compete, the quality of monetary services constantly evolves. Fiat currency dominance will not disappear, as both the U.S. government and the FED issuing fiscal and monetary policies will continue to exist. In such a world, ever-evolving sound money protocols will provide a window through which citizens can protect themselves from the side effects of recklessly issued fiat currencies.

### E.  What Are Good Properties to Inherit from Bitcoin?

(**Simplicity in consensus**) The consensus among many p2p nodes is made every 10 min over the best-effort service and often-times hostile internet. How can robust operations become possible? It is interesting to focus on the Bitcoin consensus. The nodes in the Bitcoin network are not divided according to their job. Each node performs the necessary work, confirms transactions, groups them into a block, adds the proof of work, and publishes the block as quickly as feasible. Consensus is reached as the result of each node simply performing its work for its own benefit. No node is forced to work in a time-division schedule. Each node does not need to adapt to the progress of the other nodes (therefore requiring no contact with each other) to obtain a consensus. Every node creates blocks, and every node validates the blocks. The protocol is simple and plain. This simplicity results in a robust performance. Hence, blocks cannot but be kept on produced. Rewards are provided as incentives to nodes. The more effort a node makes the more opportunities it will earn rewards. Node righteousness is not required, and no punishment that exists in a PoS-based consensus algorithm [14] is required.

(**BA algorithms are not decentralized.**) Jobs are divided under a BA algorithm. A set of nodes under the proposer's name creates blocks. The proposed blocks are validated by another group of nodes, such as *attesters*. They vote for each candidate block. Cast votes are collected and counted. A block with a satisfying number (i.e., supermajority) of votes was selected and connected to the *status quo* chain. To complete this, each node must fit into a tight schedule. Consequently, the time axis must be separated into slots and epochs. The number of nodes engaged in consensus must be limited to a few hundred to operate properly across a hostile environment such as the internet.

(**Time-energy-borne wealth: Bitcoin is a stored wealth transformed from time and energy.**) Each block header contains information on time and energy consumption. Take any blocks from the past. The block header contains the time stamp when the block was created. The difficulty level of the puzzle is also specified. Using this information, one can calculate the amount of computation (hash cycles) required to solve the hash puzzle. It would have taken more than three years for a single node working alone to solve the hash challenge. But it takes only 10 min for the entire network of operating nodes distributed independently worldwide. See the alone impossible together possible (Al-Im-To-Po) theory [33]. Thus, it indicates that hundreds of millions of computing nodes have to work together at that moment to produce that block. Given the energy efficiency of a mining node, the amount of total energy expended to create a block can be calculated. A predetermined number of new bitcoins are minted on the block. It can be said that the blockchain network has transformed the energy spent on creating the block into the Bitcoins issued from that block. The miners invested time and energy, thus are deserved to get the Bitcoins reward.

In this approach, each block generation process may be considered a new way of storing wealth. It is a transformation of the most basic resources of time and energy. Time and energy are the most valuable resources humans have and are the most fundamental types of wealth. Thus, the coins created in each block are assigned a monetary value from birth. The effort and time are not squandered; they were converted into something valuable: Bitcoin. Others have used their time and energy to create important commodities and services such as food and building houses. One can give Bitcoin to purchase goods and services that others have produced and offered.

### F.  What Are the Issues That Require Attention?

There are two problems. The first is the security risk in cryptography used in Bitcoin owing to advances in quantum computing technology [1][17]. The second is the huge energy consumption issue of the Bitcoin network.

As mentioned earlier, IBM is close to building a quantum computer that is known to break well-established cryptographic algorithms, such as elliptic curve cryptography, digital signature algorithm, and RSA algorithm. President Biden has signed two executive orders [50]. The first is to address how US leadership can be maintained in quantum computing technology. The second is to outline the U.S. government's strategy for mitigating the risks to vulnerable cryptographic systems due to advances in quantum technology. We address this further in Section V.

Bitcoin's energy consumption is enormous; its annual energy consumption has grown to 138 TWh in early 2022, which is more than the power consumption of a country such as Norway. Its annual carbon dioxide emission reached 114 million tons, which is comparable to that of Belgium [49][54].

Bitcoin advocates claim:

- The energy usage is still a very tiny proportion of global electricity consumption.
- Miners with self-interest move to places where surplus gas, solar panel farms, hydro and wind power plants, low-carbon nuclear plants, and geothermal energy are available.

But many still express concerns [50][54]:

- Chinese ban on Bitcoin mining was a response to a power deficit.
- Kazakhstan has imposed a limit on it owing to energy shortages.
- Sweden has called for the Europe-wide ban, blaming it for slowing climate transition.
- Tesla has withdrawn from its plan to accept Bitcoin as a form of payment, citing environmental concerns.

### G. Alternatives miss the merits of Bitcoin

Detractors claim that energy-intensive PoWs are the culprit. Rival currencies that use variants based on proof-of-stake (PoS), such as Ethereum and Solana, are gaining popularity. Ethereum [14] announced its plan to reduce its estimated energy consumption by up to 99% with the completion of the transition to its PoS variant system [5][7][8][14]. Ethereum completed the Merge on September 15, 2022.

However, PoS is known to have many obvious concerns [7][10][35]. The PoS is being introduced to address energy concerns and increase the transactions per second (TPS). However, the penalty may be significant because it is neither decentralized nor secure. The PoS is not a technological advancement; it lacks the advantages of the PoW. It resorts to a sociopolitical solution: Plutocratic politics and the time-energy-born wealth property is lost. Because the PoS does not retain any energy on a block, the blocks may be readily rewritten; hence, it is insecure. Advocates of PoS put forward a policy for bad actors, such as "your stakes will be confiscated if you act badly," This is a "fixing a barn after losing a cow" approach. The richest few can make confidential agreements off-chain and take control of the blockchain. These off-chain conspiring operations leave no on-chain trace; thus, no one can become aware of them. Hence, it is sensitive to bribery and conspiracy [38]. There is a nothing-at-stake risk [35]. There is a risk of a grinding attack if the random function for selecting a block creation node is unfair or predictable [10].

A PoS-based alternative may serve as a global computing platform but may not be decentralized, nor secure and thus not suitable for a global monetary grade network such as Bitcoin.

## IV. GREEN BITCOIN

In this section, we discuss Green Bitcoin.

### A. Key Performance Metrics

A set of key performance metrics of Green Bitcoin are:

M1. PQ secure cryptography and consensus
M2. Energy consumption efficiency (ECE)
M3. Byzantine fault tolerance (1/2)
M4. The mining schedule (21 million Bitcoin)
M5. The block generation time (10 min average), and
M6: The block size (1 mb)

Note that M4, M5, and M6 are the same as those of Bitcoin. M1 and M2 are our major focus. We aim to develop a PQ secure cryptography and a new energy-efficient consensus mechanism. For M3, as later shown in this paper, Green Bitcoin supports a BFT of up to 50%.

### B. Green Bitcoin Consensus

The Green Bitcoin consensus comprises two new major parts: a verifiable (self-election) coin-toss function (VCT) and a novel proof-of-computation (PoC) primitive. A simple PoolResistantComp algorithm can be used to aid these two parts. Green Bitcoin will base its PoC part on a recently published anti-ASIC technology known as the error-correction code proof of work (ECCPoW) [30][31][45]. A more detailed discussion on ECCPoW is provided in Section V.F. A critical component of the virtual machine will also be enhanced; in particular, elliptic curve cryptography will be replaced by our new PQ safe cryptography.

We aim to address the PQ security and environmental concerns while maintaining decentralization. We will let computational challenge decreased. More nodes will be able to join the network as computational difficulty decreases. As a result, security will be bolstered via enhanced decentralization. The time-energy-borne wealth property of Bitcoin will be left de-emphasized. We aim to achieve the sound money property without its price relative to fiat soaring.

Such goals will be met through the development of new on-chain means. Each node performs a simple task independently; the system's simplicity allows the system to run steadily even with a large body of participant nodes.

The consensus protocol is simple. Each participating node can easily obey the rules, and each node performs the same simple job. No job and time are divided. They need no inter-node communications to reach a consensus; each node performs its work independently. The same procedure is repeated for each new block; consensus for the current block is completed when some nodes announce the valid next block. The only announcement that each participant needs to keep a vigil for is the announcement of this valid new block. This simplicity lowers the entry barrier and invites more participating nodes.

Thus, Green Bitcoin enables the construction of a decentralized, scalable, and secure consensus solution for an extremely large network of participating nodes. It works well even if the number of participating p2p nodes exceeds one million.

(**The base-set of p2p nodes**) The base set of p2p nodes is defined as all the nodes participating in transaction validation and block formation. The protocol is set to perform well with base-set sizes greater than one million.

(**Green Bitcoin consensus**) Like the hash PoW, all nodes in the base set collaborate and contribute to creating each new block. Each node self-selects as a validator, attaches a proof for it, validates transactions, forms a block, and attaches a proof of the solution. Each node repeats this procedure for each new block. The benefit of this is the simplicity of the algorithm. Finality is determined by the longest chain or the amount of energy stored (measurable by each block difficulty level) in the blockchain. If there are two blockchains, each node for its own benefit will select the one with the most energy stored and connect a new block into it.

(**Verifiable coin-toss function**) Let us consider a coin toss game. Every node has its own unique (secret key) coin. Each node tosses its coin. A coin toss features a single output that may be either a pass or fail. Verifiable coin-toss function (VCT) is a verifiable random function (VRF). It has two inputs. One input is the secret key of the node, and the other is the previous block header. Thus, each node cannot but toss this VCT once and only once for each block. The purpose of VCT is to provide a means to turn off a certain portion of the base-set nodes, thus saving energy while allowing a large number of nodes to participate. If the probability of failure is set to 90%, 90% of network nodes are put to rest, and thus energy savings of 90% is achieved. To work on a new next block, each node tosses coin again.

We attempt to design the VCT function so that the odd of pass can be controlled. The probability of pass is a critical parameter that the network designer can use to change the amount of energy saving, given the size of the base set. For example, when the number of nodes participating in the network is small, it can be set to 100% to maximize security.

For those nodes that have been self-selected to perform the computing work, there are three types of computations: VeriComp, SolComp, and PoolResistantComp, each of which has to be performed.

(**VeriComp**) validates transactions and compiles them into a new block. Each node sets out to validate the new block upon receiving a new block announcement. If the block is valid, the node sets out to begin extending a new block next to the validated block.

(**SolComp**) is the computation required to solve the crypto puzzle. Each round presents a completely fresh crypto-puzzle. The puzzle problem is not predictable in advance, but it is determined if the preceding block header is fixed. Each node in the self-elected set starts the race to solve the crypto-puzzle as quickly as possible. A node with a proof of solution inserts the proof into the block header and broadcasts the new block instantly.

(**Coin is tossed before any energy is spent on SolComp.**) In the Green Bitcoin consensus, we observed that all nodes participate in the same simple routine. Each node performs a random turn to form a new block and attaches the proof of solution. The new VCT function makes it possible for them to take random turns. Each node progresses into the energy hungry SolComp routine if its coin toss result is a pass.

### C. What If Each Selected Node Forms a Mining Pool?

It is possible that the nodes selected by tossing coins can promise a reward distribution to the unselected nodes and request collaboration. Selected nodes would enjoy such collaboration because they can use them to stay ahead in competition among the selected nodes and increase the probability of winning the block reward. Unless there is a deterrent to prevent selected nodes from making these choices, the proposition that VCT can reduce energy consumption may turn out to be untrue.

The key question then is: Can we find an on-chain means to discourage such cooperation? Can we find a way to reduce the incentive for a selected node to form a pool by itself? Such a concern has its roots in pool mining practice in current PoW mining networks such as Bitcoin and Ethereum Classic. In a mining pool, miners subscribe to a mining pool server, perform mining tasks, and share rewards by submitting proper solutions.

To quickly grasp the idea, we can delineate our proposed solution for the pool-mining case. Here, the pool mining server is compared with one of the *selected* nodes in the Green Bitcoin consensus mechanism; the miners subscribing to the pool mining server can be compared with the *unselected* nodes. Each selected node can solicit cooperation from unselected nodes and increase its probability of winning the block reward. Unselected nodes can increase their chance of winning a share of the block reward by helping the selected node. Because both parties may benefit from such cooperation, unselected nodes continue to spend energy doing the SolComp work even if they were unselected in the first place by coin-toss. This would erase the energy reduction effect. In a pool mining protocol [46], five major steps are required for miners to obtain shared rewards:

1) register itself as a miner to a Stratum server
2) get the block header information
3) work until a mining success is announced
4) submit the mining results, and
5) get shared rewards from the server.

The key idea in breaking apart the cooperation between the selected and the unselected nodes is to create a conflict between them. One such scheme can be designed such that the selected node is asked to have its private key at risk revealed in order to delegate SolComp work to the unselected nodes.

(PoolResistantComp) There is a method proposed in the literature [15] that aims to break the cooperative tie between the pool server (vis-a-vis the selected node in our case) and the miners (the unselected nodes), thus discouraging pool mining. This is known as a two-phase PoW system. The first PoW is the routine PoW part, which is the same as the existing Bitcoin crypto-puzzle. Miners seek to find the hash of the header that is smaller than the published difficulty parameter. The second PoW is PoolResistantComp: it is the second puzzle in which the miner is asked to include a critical piece of information as an input to the hash function. Namely, the miner is asked to have the header signed with the private key of the coin-base transaction, that is, SHA256 (sign (header, privkey)), and the second puzzle is resolved if a node successfully presents a hash of that signature that is smaller than a second difficulty parameter.

Note here that the second puzzle cannot be relegated to unselected nodes unless each selected node takes the risk of revealing its private key to the unselected nodes.

In our case, we do not need such a two-phase protocol. The PoolResistantComp routine can be incorporated right into the SolComp stage with minimal effort. The PoolResistantComp part is a routine asking for the inclusion of the private key corresponding to the coin-base transaction. ECCPOW (see Fig. 1 in Ref. [45]) is a routine in which a decoder finds a codeword from the hash output, that is, OUT = SHA256(current block header). To embed PoolResistantComp into it, we can replace the hash output routine to include the private key's signature, i.e., OUT = SHA256(sign(current block header, privkey)). Such a single-line update is sufficient to regulate the nodes and force them to rest if they are not selected. This update does not lose the characteristics of *the simple* and *plain* routine discussed in IV.B. Each node simply does its routine, regardless of the states of the other nodes.

## V.   NOVEL PQ SECURE PRIMITIVES

We now discuss the novel postquantum-ready and suitable cryptographic primitives. They are the novel key generation, sign, and verification functions; it also has a new Green Bitcoin, VRF, VCT, and VC.

Quantum computers are known to break well-established cryptographic algorithms such as the elliptic curve cryptography, digital signature algorithm, and RSA algorithm. In particular, these methods are based on integer factorization and discrete logarithm problems, which are known not quantum-safe. In contrast, code-based cryptography issues are known to be quantum-safe.

(**A brief history of early code-based cryptography**) McEliece first presented code-based cryptosystems using binary Goppa codes in 1978 [37]. In 1986, Niederreiter proposed a knapsack-type public-key cryptosystem based on error-correction codes using GRS codes [42]. Subsequently, the Niederreiter method was demonstrated to be as secure as the McElice cryptosystem. Sidel'nikov and Shestakov demonstrated in 1992 that Niederreiter's plan to employ GRS codes was insecure [48]. Various methods have been proposed to minimize the public key size by employing different codes, such as the Gabidulin code [18][19], algebraic geometry code [20][28], and Reed-Muller code [47]. However, all of these approaches ultimately proved to be unstable [34][44].

### A. Recent PQ Secure Signature Primitives

This section covers recent advances in PQ cryptography and selects a set of suitable PQ secure algorithms. They can be used to meet our goal of developing a PQ secure signature and a PQ secure VRF for Green Bitcoin.

A digital signature (DS) algorithm comprises three parts. The first is the KeyGen component, which produces a public and private key pair. The second is the Sign part. Given the message and private key, it generates a signature. The third is the Verify component, which generates a binary pass or fail output based on a message and signature.

A VRF is similar to the DS algorithm in that it comprises three functions: a KeyGen function that generates a private and public key pair, a VRF function that outputs a signature (proof), and a random number, given the input of a private key and a message, and a Verify function that generates an output of pass or fail, given the input of the public key, message, random number, and proof.

We conducted preliminary research and discovered that Dilithium [36], Falcon [52], and Durandal [2] are good candidates for PQ-safe signature algorithms suitable for Green Bitcoin: the key metrics we used to select them are the size of the keys, the size of the signatures, the time it takes to complete a sign, and the time it takes to verify.

Let us now compare them. The time unit is msec. Dilithium requires 1.4, 6.2, and 1.5 for KeyGen, Sign, and Verify, respectively. Falcon, however, was 197.8, 38.1, and 0.5. Durandal received four, four, and five points, respectively. Durandal, therefore, has the quickest signature time compared with rival methods. The signature time corresponds to the VRF generation time. Thus, Durandal could serve as the first candidate for building a fast PQ-safe VRF. It offers a code-based DS algorithm for ranking metrics. While rendering a quantum-safe signature, it is sufficiently concise. The signature is 4 kb (kilo byte) long, whereas the public key is approximately 20 kb. The signature and verification processes took only 4 ms and 5 ms, respectively. It is powerful, quick, and concise enough to be considered a candidate for a worldwide public monetary-grade blockchain, such as Green Bitcoin.

We proceed to carefully study these candidate PQ secure algorithms and select the best one that satisfies all the key performance metrics of Green Bitcoin. The selected signature algorithm will be implemented using C++, and we will replace elliptic curve DS cryptography.

### B. Novel PQ Secure VRF, and VCT Functions

We aimed to discuss how to create a novel PQ secure VCT function. To this end, we first need to create a good PQ secure VRF function. We then use it to create the VCT function.

A VRF is a function that generates a unique random number with a unique signature (proof) attached to it, given a private key and message. It is distinguished from an ordinary random number generator because it also offers a verification procedure. Thus, any verifier can check whether the random number is properly calculated. The quality of the random number generated must be high; given the size of the keys, the entropy of the random number is maximized.

A VRF is similar to a DS scheme, as mentioned in Section V.A. However, a significant difference was observed. It is the signature. The signature for a DS method needs to be non-unique stochastic by design. Stochastic signatures can increase security. However, for a VRF, the signature must be unique to each fixed input. We recall our goal of using VRF to save energy in the SolComp stage. Hence, the VRF should be designed to execute just once and only once every block; otherwise, the node would abuse it by running the VRF as many times as possible until the node produces a suitable output; no energy savings are realized as a result. Such enforcement is achieved if VRF generates a unique signature for a given fixed input message.

The input can be designed to be a piece of public information that existed before the VRF was performed, such as the block header of previous block. The same applies to the public keys. The public key should have been already posted somewhere in the blockchain before running the VRF. Therefore, the private key associated to the fixed public key is fixated. As the result, each node cannot but run it once and only once per each block.

### C. Our Approach

Creating a new routine within a cryptographic algorithm is generally difficult because it changes the method's security output. However, it appears that there are a few options for this scenario. Any modification to make a unique proof reduces the degree of freedom (DoF) in the proof part. The system can remain secure if the same amount of DoF is increased elsewhere. To illustrate, suppose that the DoF in the proof is reduced so that the proof is made unique for a given fixed input. We now aim to discuss two possible directions. First, we add the same degree of DoF to the random value; the VCT function may be made to accommodate this modification, which slightly increases the size of the random value. Second, we may take the approach of increasing the same degree of DoF in the private key and this will slightly increase the size of the private key. A right balance shall be found to ensure that such a change does not compromise security.

### D. Cryptanalysis for PQ secure VRF

The security of the proposed VRF can be evaluated using routine methods [3][4] such as (a) uniqueness analysis, (b) collision resistance analysis, and (c) pseudo-randomness analysis. Our proposed VRF candidate's performance can be measured by analyzing the time required for making the proposed signature, followed by the hashing time and the overall time, which includes key generation, sign, proof, verification, and building a block [13][51]. Other factors, such as the size of private and public keys, length of signature and hash, computational complexity, and energy consumption, can all be included to evaluate efficiency [16].

### E. How to Make a VCT function from the VRF Output?

Given a new good VRF function, we aim to utilize it to create the VTC function. The VCT function accepts the result of the VRF function as input and produces a binary output, pass or fail. Namely, the output space of VRF can be partitioned into two. VRF output is a number. To create a VCT, all we need is a threshold. If VRF output is a number larger than or equal to the threshold, then VCT is set to give a pass output. The probability of pass can then be controlled by raising or lowering the threshold. The probability of pass can be set for any particular energy efficiency goal.

In addition, the probability of pass can be weighted based on the option the network designer can choose. For example, the designer can choose to enable a PoS option. A node's probability of passing can be determined by the stake it has made on chain.

### F. PQ Safe Error-Correction Codes PoW

For the VC part, we aim to use ECCPoW for its ASIC resistance, simplistic, time-varying, PQ ready properties. ECCPoW is a new VC method the first author has developed and published [30][31][45]. It is based on an error-correction code called low-density parity check (LDPC) code. It works like an error-correction code (ECC) based cryptosystem. The resistance characteristic of ECC cryptosystems to quantum Fourier sampling attacks has been demonstrated [11]. Faster and more secure ECC cryptosystems are still under study [1]. Durandal, for example, is a lightweight and secure rank-metric code-based cryptosystem [2].

In summary, we aim to extend ECCPoW in two ways. The first is to make it a Green Bitcoin suitable (see M1–M6). The second is to make it PQ safer using medium-density codes. We also aim to ensure that these extensions are safe from well-known security threats.

(**Difficulty Control Algorithm**) The tradeoff relationship between the amount of verifiable computation and energy expenditure can be precisely determined. This tradeoff relationship can be used to devise a difficulty control (DC) algorithm. The DC algorithm seeks to create blocks in a defined regular (average) interval while reacting to variation in the total number of participating p2p nodes over time.

Here, we discuss a way to make ECCPoW PQ safer. In ECCPoW, LDPC codes are used to generate time-varying crypto puzzles. We aim to replace it with moderate-density parity check (MDPC) codes and make ECCPoW PQ safer. LDPC codes do not possess any algebraic structure but only a simple combinatorial property, i.e., sparsity in the parity-check matrix; this makes them postquantum secure. There have been various suggestions to make a McEliece scheme using LDPC codes [3][4][39][40]. The low-weight parity-check rows in the party-check matrix correspond to low-weight codewords in dual codes. As such, sparsity can be utilized to draw cryptographic attacks. Consequently, MDPC codes have been proposed in which the density is increased about ten times. Furthermore, a quasi-cyclic structure was devised for shorter public and private keys [43]. One famous example of this cryptosystem is BIKE, one of the third-round algorithms in NIST Post-Quantum cryptography (PQC) standardization [1].

### G. BFT and Energy Consumption Efficiency of Green Bitcoin

Each node runs the VCT, sees the outcome either as pass or fail, and advances itself to the verifiable computation stage if it is passed. Green Bitcoin, therefore, supports a BFT of 1/2. Suppose a base set of a certain size for Green Bitcoin nodes. To launch a 51% double-spending attack, the attacker must hold 51% of the seats on the SolComp Committee. Using VCT only decreases the overall size of the committee but does not affect its proportion. This necessitates the adversary to have 51% presence in the base set to launch a 51% attack. This remains valid regardless of VCT's pass probability.

Green Bitcoin can choose a certain pass probability (PP) to determine its network's energy consumption efficiency (ECE). An ECE of 90% can be achieved when PP is set to 10%.

## VI. TESTBED AND DISCUSSION

### A. Testbed

The Green Bitcoin protocol suite will be developed on an existing open-source Bitcoin suite. The opcode table will be upgraded with the developed Green Bitcoin cryptography, replacing elliptic curve encryption for sign and verification. We will present emulation results on a proof-of-concept (PoC) network with a larger number of p2p computers. The Amazon Web Services will be used. Nodes will be scattered around the globe. We aim to employ more than a thousand nodes. Quantum attacks [18] will be used to assess the security of the PoC network.

Similar to our ECCPoW implementation [30][45], we will use the C++ as the developer language. A new genesis block will be created. To build a client, Green Bitcoin will be selected as the consensus algorithm. After defining the chain ID and exporting the genesis, the test Green Bitcoin network will be launched.

### B. Safeguarding against Profitable Double-Spending Attacks

The most secure PoW protocols are still vulnerable. Double-spending attacks are still possible to occur [27][33]. This problem is exacerbated when the network's computational power is small. By borrowing computational resources from a mining rig lending site, an attacker can launch a DS assault. If there is a profit-taking opportunity, an attack is possible. Such an opportunity opens up as long as profits overwhelm the costs. The key new finding in [28] is that profitable double-spending (PDS) attacks can occur even if honest nodes have more than 50% of the computational resources of the network. The attacker can attempt to double-spend a transaction whose stake exceeds the cost of leasing mining machines from a lending service. Green Bitcoin aims to safeguard networks against PDS attacks. Such assaults cannot be completely forbidden, but they can be discouraged by lowering the profit the attacker can make and increasing the cost the attacker must bear. Micropayments are not affected by this, but large transactions require attention. We will develop new APIs based on [28] and use them to secure large transactions. These will be made available to the global research community.

### C. Discussion

Bitcoin improves self-sovereignty of individuals. One can move along with one's stored wealth in Bitcoin anywhere in the world. No powerful entity can confiscate the portable wealth stored in Bitcoins. One can memorize the pass praise, move to a new country, and restore one's wallet. One does not need to worry about bandits on the travel route or a government's confiscation at the entry point of a port. One can make international payments anywhere in the world using Bitcoin. Currently, Bitcoin faces energy concerns. Green Bitcoin resolves this with its VCT and PoolResistantComp mechanism. Bitcoin faces quantum computer risk. Green Bitcoin provides PQ secure computations. Green Bitcoin is a new protocol designed to retain the merits of Bitcoin and addresses the two pressing concerns of Bitcoin.

# VII. CONCLUSION

Imagine a society where sound money is restored and prevails; suppose there is no cheap money. In such a society, when a man has money in his savings account, it means that he has done something beneficial to others in the past. Because this was the only way for him to earn sound money. He earned it. He was paid by the payer. It means he satisfied the payer. To satisfy the payer, he must have worked hard to produce useful goods and services. He reached his silver age. He no longer can work and produce. But he has kept sound money in his wallet. It has retained good value because it is a sound money. He can use it to purchase the valuable goods and services he needs. He does not need a government for that. He could purchase a house in a nice neighborhood and put food on the table for his family. Unearned income disappears in a world dominated by sound money, and honest work prevails.

We made two technical proposals to enhance Bitcoin. The new bitcoin (Green Bitcoin) addresses two issues: reduction of energy consumption and post-quantum computer risk. In the future, we plan to complete the Green Bitcoin protocol and bring forth a Green Bitcoin network to life. Green Bitcoin is a proposal to upgrade Bitcoin, the best sound cryptocurrency, into its quantum-safe and energy-efficient version.

**Heung-No Lee** received the B.S., M.S., and Ph.D. degrees from the University of California, Los Angeles, CA, USA, in 1993, 1994, and 1999, respectively, all in electrical engineering. He worked at HRL Laboratories, LLC, Malibu, CA, USA, from 1999 to 2002 as a Research Staff Member. From 2002 to 2008, he was Assistant Professor with the University of Pittsburgh, PA, USA. Since 2009, he has been with the School of Electrical Engineering and Computer Science, GIST, South Korea, where he is a full professor. His technical works are in information theory, signal processing theory, communications/networking theory, and their applications to Wireless Communications, Networking, Medical Imaging, Brain-Computer Interfaces, Spectroscopy, Cryptocurrencies and Decentralized Finance.

**Young-Sik Kim** received the B.S., M.S., and Ph.D. degrees in electrical engineering and computer science from Seoul National University, in 2001, 2003, and 2007, respectively. He joined the Semiconductor Division, Samsung Electronics, where he worked in the research and development of security hardware IPs for various embedded systems, including modular exponentiation hardware accelerator (called Tornado 2MX2) for RSA and elliptic-curve cryptography in smartcard products and mobile application processors, until 2010. He is currently a professor with Chosun University, Gwangju, South Korea. He is also a submitter for two candidate algorithms (McNie and pqsigRM) in the first round for the NIST Post Quantum Cryptography Standardization and a submitter for three candidate algorithms for Korean Post-Quantum Cryptography Competition (KpqC).

**Dilbag Singh** received a Ph.D. degree in computer science and engineering from the Thapar Institute of Engineering and Technology, Patiala, India, in 2019. He was an Assistant Professor at Chandigarh University, Mohali, India; Manipal University Jaipur, Jaipur, India; and Bennett University, Greater Noida, India. In 2021, he joined the School of Electrical Engineering and Computer Science, Gwangju Institute of Science and Technology, Gwangju, South Korea, where he is currently affiliated. His research interests include image processing, computer vision, deep learning, metaheuristic techniques, and information security. He was in the top 2% list issues by "World Ranking of Top 2% Scientists" in 2021.

**Manjit Kaur** received the M.E. degree in information technology from Punjab University, Chandigarh, India, in 2011, and the Ph.D. degree from the Thapar Institute of Engineering and Technology, Patiala, India, in 2019. She was an Assistant Professor at Chandigarh University, Mohali, India; Manipal University Jaipur, Jaipur, India; and Bennett University, Greater Noida, India. In 2021, she joined the School of Electrical Engineering and Computer Science, Gwangju Institute of Science and Technology, Gwangju, South Korea, where she is currently affiliated. Her research interests include wireless sensor networks, digital image processing, and metaheuristic techniques. She was in the top 2% list issues by "World Ranking of Top 2% Scientists" in 2021.